\documentclass[aps,showpacs,prd,twocolumn,superscriptaddress,floatfix]{revtex4}
\usepackage{amsmath, amsfonts, hyperref}
\usepackage{graphicx}
\newcommand{\beq}{\begin{equation}}
\newcommand{\eeq}{\end{equation}}
\newcommand{\beqn}{\begin{eqnarray}}
\newcommand{\eeqn}{\end{eqnarray}}

\newcommand{\mnras}{Mon.\ Not.\ R.\ Astro.\ Soc.}

\begin{document}
\title
{Weak annihilation cusp inside the dark matter spike 
about a black hole}
\date{\today}
\author{Stuart~L. Shapiro}
\altaffiliation{Also Department of Astronomy and NCSA, University of
  Illinois at Urbana-Champaign, Urbana, IL 61801}
\author{Jessie Shelton}
\affiliation{Department of Physics, University of Illinois at Urbana-Champaign, Urbana, IL 61801}

\begin{abstract}
  We reinvestigate the effect of annihilations on the distribution of
  collisionless dark matter (DM) in a spherical density spike around a
  massive black hole. We first construct a very simple, pedagogic,
  analytic model for an isotropic phase space distribution function
  that accounts for annihilation and reproduces the ``weak cusp''
  found by Vasiliev for DM deep within the spike and away from its
  boundaries. The DM density in the cusp varies as $r^{-1/2}$ for
  $s$-wave annihilation, where $r$ is the distance from the central
  black hole, and is not a flat ``plateau'' profile. We then extend
  this model by incorporating a loss cone that accounts for the
  capture of DM particles by the hole.  The loss cone is implemented
  by a boundary condition that removes capture orbits, resulting in an
  anisotropic distribution function.  Finally, we evolve an initial
  spike distribution function by integrating the Boltzmann equation to
  show how the weak cusp grows and its density decreases with time. We
  treat two cases, one for $s$-wave and the other for $p$-wave DM
  annihilation, adopting parameters characteristic of the Milky Way
  nuclear core and typical WIMP models for DM. The cusp density
  profile for $p$-wave annihilation is weaker, varying like $\sim
  r^{-0.34}$, but is still not a flat plateau.
\end{abstract}

\maketitle

\section{Introduction} 

A supermassive black hole (SMBH) will steepen the density profile of
dark matter (DM) within the hole's sphere of influence, $r_h =
GM/v^2_0$.  Here, $M$ is the mass of the hole and $v_0$ is the (1D)
velocity dispersion in the innermost halo just outside $r_h$. The
precise profile for this DM density spike depends both on the
properties of DM and the formation history of the SMBH. If the DM is
collisionless with a cuspy, spherical, inner halo density obeying a
generalized Navarro-Frenk-White (NFW~\cite{NavFW97}) profile then the
density profile in the absence of the hole will follow a power-law,
$\rho(r) \sim r^{-\gamma_c}$.  Simulations with DM alone yield typical
values of $0.9 \lesssim \gamma_c \lesssim
1.2$~\cite{DieKMZMPS08,NavLSWVWJ10}, but if baryons undergo dissipative
collapse into the disk they can induce the adiabatic contraction of
the central DM halo into a steeper power
law~\cite{BluFFP86,GneKKN04,GusFS06}, with values as high as $\gamma_c
\sim 1.6$ allowed for the Milky Way~\cite{PatIB15}.

If the SMBH grows adiabatically from a smaller seed~~\cite{Pee72a} the
SMBH then modifies the profile inside $r_h$, forming a DM spike within
which $\rho(r) \sim r^{-\gamma_{\rm sp}}$, where $\gamma_{\rm sp}=
(9-2\gamma_c)/(4-\gamma_c)$~\cite{GonS99}.  For $0 < \gamma_c \leq 2$
the power-law $\gamma_{\rm sp}$ varies at most between 2.25 and 2.50
for this case. However, gravitational scattering off of a dense
stellar component inside $r_h$ could heat the DM, softening the spike
profile and ultimately driving it to a final equilibrium value of
$\gamma_{\rm sp} = 1.5$~\cite{Mer04,GneP04,MerHB07}, or even to
disruption~\cite{WanBVW15}.  Other spikes,
characterized by other power laws, are obtained from different
formation histories for the BH within its host halo, such as the
sudden formation of a SMBH through mergers or gradual growth from an
inspiraling off-center seed \cite{UllZK01}, or in the presence of DM
self-interactions \cite{ShaP14}, as reviewed in
e.g.~\cite{ForB08,FieSS14}.

DM annihilations in the innermost region of the spike weaken the
density profile there. For standard WIMP models, wherein the
annihilation cross section $\langle \sigma v\rangle$ is a constant
(i.e., $s$-wave annihilation) it was suggested~\cite{GonS99} that an
``annihilation plateau'' would form at the central region of the
spike, in which case the DM profile would be flat.  Let $r=r_{\rm
  ann}$ be the radius at which the DM density in the spike reaches
$\rho_{\rm ann}$, the ``annihilation plateau'' density.  At this
radius the annihilation time scale equals the Galaxy age $T$, so that
\begin{equation}
\label{rhoann}
\rho_{\rm ann} = \frac{m_{\chi}}{\langle \sigma v\rangle T} 
\end{equation}
Here $m_{\chi}$ is the DM particle mass.

Vasiliev~\cite{Vas7} subsequently showed that an annihilation plateau
arises only if all DM particles move in {\it strictly circular orbits}
about the central black hole.  He demonstrated that if the DM
distribution function is {\it isotropic}, which he noted was 
likely, the density continues to rise with decreasing distance $r$
from the black hole, forming a ``weak cusp'' and not a plateau. Within
the weak cusp the density increases as $r^{-1/2}$ for $s$-wave
annihilation.  The reason is that particles in eccentric orbits with
apocenters outside $r_{\rm ann}$ continue to contribute to the density
inside $r_{\rm ann}$ and thereby maintain a weak inner cusp.

The distinction between an ``annihilation plateau" and a ``weak cusp''
may have important observational consequences. Due to
their extraordinarily high DM densities, BH-induced density spikes can
appear as very bright gamma-ray point sources in models of
annihilating DM
\cite{GonS99,Mer04,GneP04,GonPQ14,FieSS14,BelS14,LacBS15, SheSF15}.
Many of these models are now becoming detectable with the current and
near-future high-energy gamma ray experiments, and indeed the
excess of $\sim 1-5$ GeV gamma rays from the inner few degrees of the
Galactic Center (GC) observed by {\it Fermi} may be prove to be a
first signal of annihilating DM~\cite{DayFHLPRS16,CalCW15,FermiGCE16},
although tension with limits from dwarf galaxies \cite{FermiDwarf15}
and the statistical properties of the photons in the GC excess,
\cite{LeeLSSX16,BarKW16} may indicate an astrophysical explanation for
the GC excess such as a new population of pulsars (see, e.g.~,
\cite{AbaCH14, BraK15, OleKKD16}).  In any case, self-annihilating DM
within a spike can easily lead to gamma-ray point sources bright
enough to be seen potentially by existing gamma-ray
telescopes~\cite{FieSS14,SheSF15}.
 Now the dominant contribution to the
annihilation signal from the spike comes from the region near $r_{\rm
  ann}$. This holds whether it originates from DM $s$-wave or from
$p$-wave annihilations~\cite{FieSS14,SheSF15}.  The magnitude of the
signal thus depends on the density and velocity profiles in the region
where the spike transitions to a weak cusp.

This result seems not be have been fully appreciated, since it has not
been incorporated in many recent applications. Consequently it seems
worthwhile to revisit the issue. In general, a weak cusp of 
this form is obtained
whenever DM initially following a power-law density profile attains
sufficiently high densities that its self-annihilation becomes
important.  Thus in principle a weak cusp can form even in the absence
of a spike, e.g. for a standard NFW cusp, $\gamma_c=1$.  In practice,
given typical Galactic parameters and a thermal $s$-wave annihilation
cross section, the DM density would only reach $\rho_{\mathrm{ann}}$
for radii very near the BH, rendering the weak cusp observationally
insignificant.  For $p$-wave annihilations, and for $\gamma_c \lesssim 1$, the
weak cusp would not exist at all in this case.

We begin by providing a simple physical argument leading to analytic
expressions for an isotropic phase space distribution function and
resulting density and velocity dispersion profiles in a DM spike with
a weak cusp. Our radial density profile for this case agrees with the
result found by Vasiliev~\cite{Vas7}, who provided a scaling argument
that also allows for an anisotropic initial spike.  We next refine our
analytic model by incorporating a loss cone boundary condition that
accounts for the direct capture of DM particles by the black hole,
making the distribution anisotropic. Finally, we integrate the
collisionless Boltzmann equation numerically, allowing for an
anisotropic distribution function, and study how the weak cusp forms
in the spike and grows with time. We again confirm the numerical
results reported in~\cite{Vas7} for $s$-wave annihilation but now we
extend the analysis to include $p$-wave annihilation, with cross
sections that vary as $\langle \sigma v\rangle \propto v^2(r)/c^2$,
where $v(r)$ is the DM velocity dispersion and $c$ the speed of light.
We find that the annihilation cusp is even weaker (i.e. less steep)
for $p$-wave than for $s$-wave annihilations, but it still is not a
flat plateau.

In Section II we present our simple, pedagogic, analytic model for an
isotropic DM spike with a weak cusp and in Section III we improve the
model by including a capture loss cone, which induces an
anisotropy. In Section IV we solve the Boltzmann equation directly and
determine the time-dependent growth of the weak cusp, both for
$s$-wave and $p$-wave DM annihilations. We adopt units with $G=1=c$
unless otherwise noted.

\section{Isotropic Model: $f = f(E)$}
\label{iso}

\subsection{Density}

An isotropic distribution function for a stationary distribution of 
collisionless matter of a single species is of the form $f=f(E)$, where $E$ 
is the
energy per unit mass of a particle. We adopt Newtonian gravitation and consider 
the energy of particles in orbit about the black hole: 
\begin{equation}
\label{E}
E = \frac{1}{2} v^2 + \Phi(r), \ \ \ \Phi(r) = -\frac{M}{r}.
\end{equation}

The mass density in the spike is obtained from the distribution function 
$f(E)$ according to
\begin{eqnarray}
\label{dens}
\rho(r) &=& 4 \pi \int v^2 f dv \nonumber \\ 
        &=& 4 \pi \int_{-M/r}^{0} [2(E+\frac{M}{r})]^{1/2} f(E) dE   
\end{eqnarray}

We will adopt the following simplification: let there be no surviving particles 
with orbits that reside entirely within $r_{\rm ann}$, while let the
particles whose orbits are either partly or entirely outside $r_{\rm ann}$
be described by the unperturbed spike distribution function. Thus we
assume that all particles that orbit entirely within $r_{\rm ann}$ have been annihilated
in the age of the Galaxy, while those which spent part or all of 
their time outside this radius have avoided annihilation altogether. Crudely, particles
spend most of their time near apocenter, not pericenter, so they are more likely to
survive whenever their orbits take them outside $r_{\rm ann}$. Mathematically,
this assumption may be expressed as
\begin{eqnarray}
\label{fE}
f&=& f(E), \ \ 0 \geq E \geq -M/r_{\rm ann}, \nonumber \\
 &=& 0,  \ \ \ \ \ \ \  E < -M/r_{\rm ann}  
\end{eqnarray}

Inserting eqn.~(\ref{fE}) into eqn.~(\ref{dens}) yields
\begin{eqnarray}
\label{dens2}
\rho(r) = 4 \pi \int_{-\frac{M}{r}}^{0} \left[2\left(E+\frac{M}{r}\right)\right]^{1/2} f(E) dE,  
                  \ r \geq r_{\rm ann}, &&\\
\label{dens3}
     = 4 \pi \int_{-\frac{M}{r_{\rm ann}}}^{0}\left[2\left(E+\frac{M}{r}\right)\right]^{1/2} f(E) dE,  
                   \ r < r_{\rm ann}. && 
\end{eqnarray}

By construction eqn.~(\ref{dens2}) gives the
unperturbed spike profile for all $r \geq r_{\rm ann}$. 
Substituting the variable $y=-Er/M$ and adopting 
a power-law spike distribution function, $f(E) = K|E|^p$, 
where $K$ is a (normalization) constant, we obtain
\begin{eqnarray}
\label{dens4}
\rho(r) &=& 2^{5/2} \pi I_{1/2}(p;1) K \left( \frac{M}{r} \right)^{(p+3/2)} 
             \nonumber \\ 
     &=& \rho_{\rm ann} \left( \frac{r_{\rm ann}}{r} \right)^{(p+3/2)},   
              \ \ \ \  r \geq r_{\rm ann},
\end{eqnarray}
where
\begin{eqnarray}
I_{1/2}(p;q)  &\equiv& \int_{0}^{q}(1-y)^{1/2}y^p dy, 
\end{eqnarray}
and $I_{1/2}(p;1) = B(p+1,3/2)$, 
where $B(x,y)$ is the familiar beta function.
For a power-law spike profile $\gamma_{\rm sp} = p+3/2$. 

\begin{figure}
\includegraphics[width=8cm]{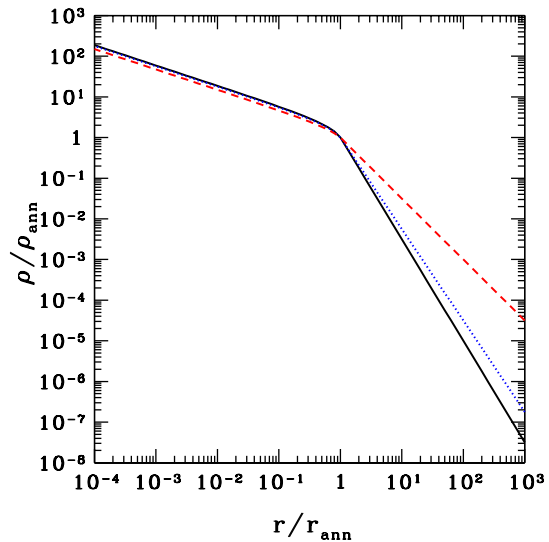}
\caption{DM density profile in an adiabatic spike around a black hole,
allowing for annihilation.
For $r \geq r_{\rm ann}$ the density varies as $r^{-\gamma_{\rm sp}}$, 
where $\gamma_{\rm sp} = 2.5$ ({\it black solid line}),
$2.25$ ({\it blue dotted line}) and $1.5$ ({\it red dashed line}). 
For $r < r_{\rm ann}$ annihilations soften the spike to 
a ``weak cusp'' with $\rho(r)  \sim r^{-1/2}$. 
Here $\rho$ and $r$ are normalized to their
values at $r_{\rm ann}$.
}
\label{fig:dens}
\end{figure}

Consider now the density profile for $r < r_{\rm ann}$ 
given by eqn.~(\ref{dens3}),
\begin{eqnarray}
\label{dens5}
\rho(r) &=& 2^{5/2} \pi I_{1/2}(p;\frac{r}{r_{\rm ann}}) K 
   \left( \frac{M}{r} \right)^{(p+3/2)} \\ 
        &=& \rho_{\rm ann} 
  \frac{I_{1/2}(p;\frac{r}{r_{\rm ann}})}{I_{1/2}(p;1)}
   \left( \frac{r_{\rm ann}}{r} \right)^{(p+3/2)},
               \  r < r_{\rm ann}. \nonumber
\end{eqnarray}
Here $I_{1/2}(p;\frac{r}{r_{\rm ann}})/I_{1/2}(p;1) = 
B(p+1,3/2;\frac{r}{r_{\rm ann}})$,
where $B(a,b;x)$ is the incomplete beta function. 
Evaluating the density for $r/r_{\rm ann} \ll 1$, noting 
$I_{1/2}(p,q) \approx q^{p+1}/(p+1)$ for $q \ll 1$, yields
\begin{eqnarray}
\label{dens6}
\rho(r) &\approx& \frac{2^{5/2} \pi}{p+1} K 
  \left( \frac{M}{r_{\rm ann}} \right)^{(p+1)} 
  \left( \frac{M}{r} \right)^{1/2} \nonumber \\
    &=& \frac{\rho_{\rm ann}}{(p+1)I_{1/2}(p;1)} 
       \left( \frac{r_{\rm ann}}{r} \right)^{1/2}, \  r \ll r_{\rm ann}.
\end{eqnarray}
Eqn~(\ref{dens6}) is exactly what we set out to prove:
the density well inside $r_{\rm ann}$ scales like $r^{-1/2}$.
Notice that this scaling behavior is independent of the 
power $p$. A continuous match between the inner and outer spike
profiles can be obtained by numerically evaluating  eqn.~(\ref{dens5})
and joining it onto eqn.~(\ref{dens4}), which we do in Fig.~\ref{fig:dens}. 

In the absence of annihilation, the adiabatic spike that
 forms in a DM cluster initially characterized by 
a power-law density profile $\rho(r) \sim r^{-\gamma}$, $0<\gamma<2$ gives
rise to a power-law profile with $2.25<\gamma_{\rm sp}<2.50$~\cite{GonS99}.
The profiles when annihilation is incorporated 
are plotted in Fig.~\ref{fig:dens} for the limiting values of 
$\gamma_{\rm sp}$.  A DM cluster that has an isothermal (and not a power-law) 
core initially forms an adiabatic spike 
with $\gamma_{\rm sp} = 1.5$~\cite{Pee72a}.
The spike profile for this value (which may also be reached if the DM spike
is subsequently heated by scattering off stars~\cite{Mer04,GneP04})
is also shown in the figure, again allowing for annihilations.

It can be shown that the contribution of DM particles
{\it unbound} to the BH, with energies $E=3v_0^2/2 > 0$, also scales
as $r^{-1/2}$ everywhere inside the BH zone of influence, i.e
$r \lesssim M/v_0^2$ 
(see~\cite{ShaT83}, eqn.~(14.2.22)). 
However, their contribution inside $r_{\rm ann}$ is much smaller in magnitude 
than the contribution of (eccentric) bound particles,
as the spike density of these contributing bound particles 
is much larger than unbound particles.

We also note that at first glance there is nothing in the above argument that
distinguishes $s$-wave from $p$-wave annihilation. The key point is
that the time scale for annihilation
decreases with decreasing $r$ in a canonical spike. It is this
feature that is reflected in equation~(\ref{fE}) for the distribution function. 
This decrease is even more rapid as $r$ decreases for $p$-wave than for 
$s$-wave annihilation, given the additional
velocity dependence in the former case. So we again expect a weak cusp to
form in the innermost region about the black hole. However, we note that
in the case of $p$-wave annihilation the annihilation density 
$\rho_{\rm ann}$ given by eqn.~(\ref{rhoann}) is not a constant
but decreases with decreasing radius. 
For purely circular orbits we would then expect that instead of a flat 
plateau density profile inside $r_{\rm ann}$ we would have a density that 
decreases as $r$ decreases. As the orbits in the cusp are dominated by
highly eccentric and not circular orbits, the cusp will not exhibit 
this decrease. However, we do anticipate that the cusp profile 
for $p$-wave annihilations will be somewhat weaker (i.e. flatter) than for 
$s$-wave annihilations due to the decreasing value of $\rho_{\rm ann}$ with
decreasing distance.  This expectation is borne out by our solution 
to the Boltzmann equation in Section IV. 

The canonical
profiles for an adiabatic spike differ considerably from those arising
in the case of self-interacting DM (SIDM), as shown in~\cite{ShaP14}.
Moreover, the effects of annihilation are washed out for SIDM, as the
the distribution function is constantly replenished inside $r_{\rm ann}$
by DM elastic scatterings. Hence there is no transition to a ``weak cusp''
inside the spike for SIDM.

Finally, we emphasize that equation~(\ref{fE}) for the distribution 
function is only approximate. The true distribution function, though
spherical,  is not strictly isotropic and is better described by
a function of the form $f(E,J)$, where $J$ is the angular momentum 
per unit mass of a DM particle. To obtain the correct function an
integration of the time-dependent Boltzmann equation with an
annihilation sink term is required to determine $f(E,J;t)$. 
Vasiliev performed such an integration
for $s$-wave annihilation. We will repeat the calculation 
in Section~\ref{evo}, incorporating a 
capture loss cone, and also do the calculation for $p$-wave annihilation.

\subsection{Velocity Dispersion}

Now consider the (3D) velocity dispersion everywhere in the spike.
It is obtained from 
\begin{eqnarray}
\label{vel}
v^2(r) &=& \frac{4 \pi}{\rho(r)} \int v^4 f dv \nonumber \\
       &=&
   \frac{4 \pi}{\rho(r)} \int_{-M/r}^{0} \left[2\left(E+\frac{M}{r}\right)\right]^{3/2} f(E) dE,   
\end{eqnarray}
which, when eqn.~(\ref{fE}) is inserted, yields
\begin{align}
\label{vel2}
v^2(r) =& \frac{2^{7/2} \pi}{\rho(r)}K I_{3/2}(p;1)
        \left(\frac{M}{r}\right)^{p+5/2},  \ \ r \geq r_{\rm ann}, \\
\label{vel3}
    =& \frac{2^{7/2} \pi}{\rho(r)}K I_{3/2}(p;\frac{r}{r_{\rm ann}})
           \left(\frac{M}{r}\right)^{p+5/2}, \ \ r < r_{\rm ann}.
\end{align}
Here
\begin{eqnarray}
I_{3/2}(p;q)  \equiv \int_{0}^{q}(1-y)^{3/2}y^p dy,
\end{eqnarray}
where $I_{3/2}(p;1)=B(p+1,5/2)$ and where
$I_{3/2}(p;r/r_{\rm ann})/I_{3/2}(p;1) = B(p+1,5/2;r/r_{\rm ann})$.
Evaluating eqs.~(\ref{vel2}) and ~(\ref{vel3}), using
eqs.~(\ref{dens4}), ~(\ref{dens5}) and ~(\ref{dens6}) for $n(r)$, yields
\begin{eqnarray}
\label{vel4}
v^2(r) &=& \frac{3}{p+5/2} \frac{M}{r}, \ \ \ \ r \geq r_{\rm ann}, \\
\label{vel5}
       &=& 2 \frac{I_{3/2}(p;r/r_{\rm ann})}{I_{1/2}(p;r/r_{\rm ann})}
       \left( \frac{M}{r} \right), \ \ \ r < r_{\rm ann}, 
\end{eqnarray}
and
\begin{eqnarray}
     v^2(r) &\approx& 2 \frac{M}{r}, \ \ \ r \ll r_{\rm ann}. 
\end{eqnarray}

\begin{figure}
\includegraphics[width=8cm]{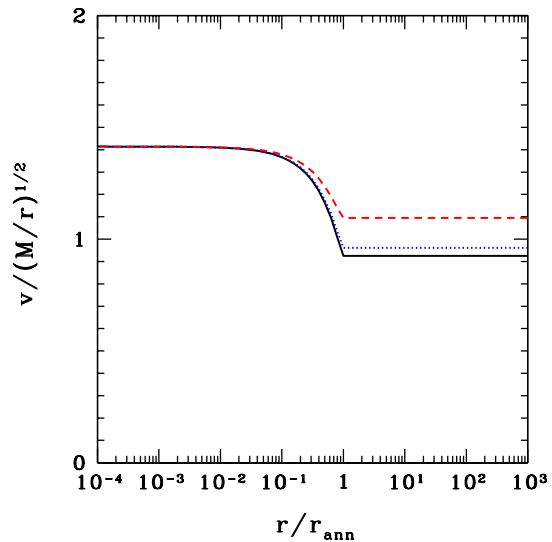}
\caption{DM velocity dispersion (3D) in an adiabatic spike 
around a black hole, allowing for annihilation. Curves are labeled as in Fig.~\ref{fig:dens}.
Here $v$ is normalized to $(M/r)^{1/2}$, where $M$ is the mass of the black hole. 
}
\label{fig:vel}
\end{figure}

The corresponding values for the 1D velocity dispersion 
$v^2_{\hat{i}}(r)= v^2(r)/3, \ \hat{i} = \{ \hat{r}, \hat{\theta}, \hat{\phi} \},$ are
\begin{eqnarray}
\label{vel6}
v^2_{\hat{i}}(r) &=& \frac{1}{p+5/2} \frac{M}{r}, \ \ \ \ r \geq r_{\rm ann}, \\
\label{vel7}
       &=& \frac{2}{3} \frac{I_{3/2}(p;r/r_{\rm ann})}{I_{1/2}(p;r/r_{\rm ann})}
       \left( \frac{M}{r} \right), \ \ \ r < r_{\rm ann}, 
\end{eqnarray}
and
\begin{eqnarray}
v^2_{\hat{i}}(r) &\approx& \frac{2}{3} \frac{M}{r}, \ \ \ r \ll r_{\rm ann}. 
\end{eqnarray}
Hence in both power-law regimes, with $\rho(r) \sim r^{-\beta}$, 
where $\beta = p+3/2$ for $r \geq r_{\rm ann}$ and $\beta = 1/2$ for 
$r \ll r_{\rm ann}$, we find $v^2_{\hat{i}}(r)= v^2(r)/3 = \frac{M}{r} \frac{1}{1+\beta}$ , as assumed in~\cite{FieSS14}.
A continuous transition between the inner
and outer spike is obtained by evaluating eqn.~(\ref{vel3}) 
numerically for $0 < r/r_{\rm ann} < 1$. We do this in  Fig.~\ref{fig:vel}
for the profiles shown in  Fig.~\ref{fig:dens}.

Finally, we note that the above results should apply well to $p$-wave
as well as $s$-wave annihilations, allowing for the smaller value of
$r_{\rm ann}$ and the slight decrease in $\beta$ in the weak cusp for
$p$-wave annihilations (from $\beta=0.5$ to $\beta \approx 0.34$; see
Section~\ref{evo}).

\section{Loss Cone: $f = f(E,J)$}
\label{lossc}

We now incorporate a realistic inner boundary condition that all
particles that {\it ever} reach inside $r_{\rm bh} = 4M$ are captured
by the black hole within a single orbital period. As a result, since
DM is assumed collisionless (except for annihilations), those capture
orbits are never replenished and the distribution function vanishes
for these trajectories.  Here we take $r_{\rm bh}$ to be the radius of
marginally bound circular orbits and the minimum periastron of all
parabolic orbits about a Schwarzschild black
hole~\cite{ShaT83,SadFW13,ShaP14}. This capture constraint induces a
loss cone in phase space: for any particle of energy $E$, there are no
particle orbits with angular momentum per unit mass satisfying $J \leq
J_{\rm loss}(E)$, where $J_{\rm loss}(E)$ is the angular momentum at
which $r_p(E,J_{\rm loss})=r_{\rm bh}$. Here $r_p(E,J)$ is the
pericenter radius of bound particles of energy $E$ and angular
momentum $J$ in (elliptical) orbit about the black hole. Accordingly,
we have
\begin{equation}
\label{Jloss}
J_{\rm loss}(E) = r_{\rm bh}\left[ 2 \left( E+\frac{M}{r_{\rm bh}} \right) \right].
\end{equation}

Following ~\cite{Vas7}, we change phase-space variables from $\{E,J\}$ to $\{E,R\}$, 
defining $R \equiv J^2/J_c^2$, where
$J_c = M/(-2E)^{1/2}$ is the angular momentum of a circular orbit of 
energy $E$. Hence $0 \leq R \leq 1$. 
Eqn.~(\ref{Jloss}) then gives
\begin{align}
\label{Rloss}
R_{\rm loss}(E) &=& 4 \frac{r_{\rm bh}}{M}(|E|(1 - |E|\frac{r_{\rm bh}}{M}),
\ \ \ 0 \geq E \ge -\frac{M}{(2 r_{\rm bh})}, 
\end{align}
Orbits with $E < -M/(2 r_{\rm bh})$ cannot avoid penetrating the inner
boundary at $r_{\rm bh}$ and hence
don't survive capture. Annihilations thus are relevant only 
when $r_{\rm ann} > 2 r_{\rm bh}$.
Incorporating the loss-cone boundary condition in our simple 
distribution function that accounts for annihilations 
when $r_{\rm ann} > 2 r_{\rm bh}$ yields 
a two-dimensional distribution function,
\begin{eqnarray}
\label{fER}
f_{\rm loss}(E,R) &=& \hspace{-0.05in} f(E),  
\ \ 0 \geq E \geq -\frac{M}{r_{\rm ann}}  ~and~  
R \geq R_{\rm loss}(E), \ \ \ \nonumber \\
  &=&  0, \ \ E < -\frac{M}{r_{\rm ann}} ~or~ R < R_{\rm loss}(E). 
\ \ \ \ \ \ \ \ \  
\end{eqnarray}
The above form guarantees that $f_{\rm loss}(E,R)=0$ for all $E  < -M/(2 r_{\rm bh})$.
Strictly a function of the integrals of motion $E$ and $J$ (or $E$ and $R$), 
$f_{\rm loss}(E,R)$ is a steady-state solution of the
collisionless Boltzmann equation, according to the Jeans Theorem.

Obtaining the density and velocity dispersion profiles generated by 
this distribution function requires a two-dimensional integration 
over velocity space inside the spike. Using the expression
\begin{equation}
d^3 v = \frac{2 \pi J_c^2 dR dE}{r^2 |v_{\hat{r}}|},
\end{equation}
where $v_{\hat{r}}$ is the radial velocity, we 
determine these moments according to
\begin{eqnarray}
\label{moments}
&&\rho(r) = 2^{-1/2} \pi \left(\frac{M}{r}\right)^{3/2} \\ 
&&\times \int^1_0  
\frac{d \varepsilon}{\varepsilon} 
\int^{4 \varepsilon (1-\varepsilon)}_0 dR ~f_{\rm loss}
(-\frac{\varepsilon}{r},R) 
\frac{1}{\sqrt{ 1 - \varepsilon - \frac{R}{4 \varepsilon}}}, 
\nonumber \\
&&\rho v^2(r) = 2^{1/2} \pi \left(\frac{M}{r}\right)^{5/2} \\
&&\times \int^1_0 \frac{d \varepsilon (1-\varepsilon)} {\varepsilon} 
\int^{4 \varepsilon (1-\varepsilon)}_0 
dR ~f_{\rm loss}(-\frac{\varepsilon}{r},R) 
\frac{1}{\sqrt{ 1 - \varepsilon - \frac{R}{4 \varepsilon}}}, 
\nonumber
\end{eqnarray}
where $\varepsilon \equiv -Er/M$.

\begin{figure}
\includegraphics[width=8cm]{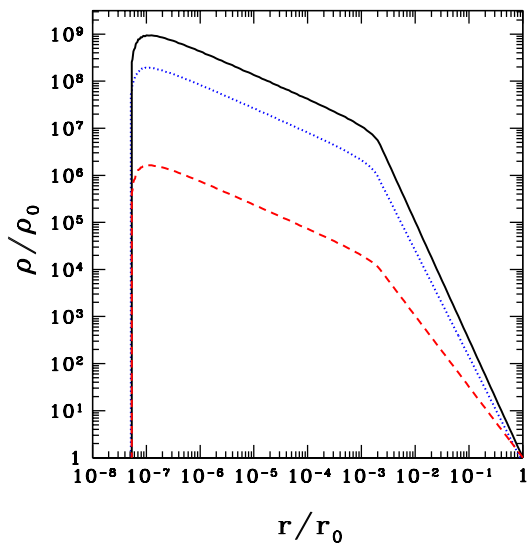}
\caption{DM density profile in an adiabatic  
DM spike around a black hole,
allowing for annihilation and black hole capture.
Curves are labeled as in Fig.~\ref{fig:dens}. The densities and
radii are normalized to their values at fiducial radius $r_0$ in the outer
spike. The annihilation radius is fixed at
 $r_{\rm ann}/r_0 = 2.2 \times 10^{-3}$
and the capture radius at  $r_{\rm bh}/r_0 = 5 \times 10^{-8}$.
}
\label{fig:dens_loss}
\end{figure}

\begin{figure}
\includegraphics[width=8cm]{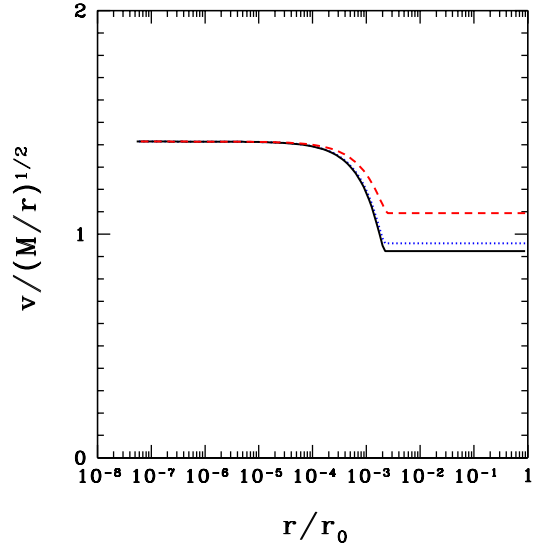}
\caption{DM velocity dispersion (3D) in an adiabatic DM spike
around a black hole, allowing for annihilation and black hole capture. 
Curves are labeled and parameters assigned as in Fig.~\ref{fig:dens_loss}.
Here $v$ is normalized to $(M/r)^{1/2}$. 
}
\label{fig:vel_loss}
\end{figure}

The results of the integrations are shown in Fig.~\ref{fig:dens_loss} for 
density profiles and Fig.~\ref{fig:vel_loss} for the velocity profiles. Shown are
curves for the same power-law spikes $f(E) = K|E|^p$ plotted 
in Figs~\ref{fig:dens} and \ref{fig:vel}, but now clipped in $R$ in accord with  
Eqn.~(\ref{fER}). Here we normalize 
radii to a fiducial outer spike radius $r_0$, where the density is assumed to be $\rho_0$.
We fix the annihilation radius at $r_{\rm ann}/r_0 = 2.2 \times 10^{-3}$
and the capture radius at  $r_{\rm bh}/r_0 = 5 \times 10^{-8}$. As we will see
in the next section, if we assign $r_0$ to reside near the outer radius of the DM spike, 
where the particles bound to the black hole join onto the ambient nuclear core, 
then these dimensionless ratios are within an order of magnitude of those inferred for
the DM spike in the Milky Way. In this case $r_0 \approx M/v_0^2$, where $v_0$ is the
velocity disperson characterizing the nuclear core and
$M$ is the mass of Sgr A*. We postpone making a more careful match to realistic 
Milky Way parameters to the next section.

As is intuitive, the above density profiles differ little from those found in
the absence of a loss-cone boundary condition, except at radii approaching $r_{\rm bh}$, 
where the loss cone grows to occupy an appreciable fraction
of phase space. For $ r_{\rm bh} \lesssim r \ll r_{\rm ann}$ we again 
find $n(r) \sim r^{-1/2}$. The magnitude of the 
3D velocity dispersion remains fairly insensitive to the 
presence of the loss cone, but the eccentric orbits, which dominate
the weak cusp, are also destroyed as $r \rightarrow r_{\rm bh}$.

\section{Evolution: $f(E,R;t)$}
\label{evo}

We now consider the evolution with time of the DM profile in the spike 
by integrating the Boltzmann equation directly, allowing for
annihilation. We adopt the approach in~\cite{Vas7}, 
but now we incorporate a loss-cone
boundary condition in $f(E,R;t)$ and treat two cases: one for $s$-wave and the
other for $p$-wave annihilation.  We again 
assume that the black hole grew to its present
mass $M$ adiabatically at the center of the inner, spherical, 
DM Galactic halo, where the density profile was $\rho(r) \sim r^{-\gamma_c}$, 
and that this growth occurred over
a time $t \ll T = 10^{10}$~yr. The result is the formation of 
a DM spike about the black hole that obeys
a new power-law density profile $\rho(r) \sim r^{-\gamma_{sp}}$,  
with $\gamma_{sp} = (9-2 \gamma_c)/(4-\gamma_c)$~\cite{GonS99},
corresponding to a power-law phase-space
distribution function $f(E) \propto |E|^p$ with $p=\gamma_{sp}-3/2$. 

We specialize to parameters 
appropriate to the Milky Way nucleus and typical WIMP particle models, 
which is the basis of the ``canonical'' adiabatic spike 
in~\cite{FieSS14,SheSF15}.
We recall that $\gamma_c = 1$ is the standard NFW value for
the central DM halo. Following~\cite{FieSS14} we take
instead $\gamma_c=1.26$, the best-fit 
value reported in~\cite{DayFHLPRS16} , which provides a recent 
analysis of the {\it Fermi} data
of the $\sim 1-3$ GeV gamma-ray excess from the Galactic center and the 
possibility that it might be a signal of 
DM annihilations. This value then yields $\gamma_{sp} = 2.36$ and $p=0.86$ for an 
adiabatic spike.

The outer boundary of the spike is taken to be at 
$r_b = 0.2r_h = 0.34$~pc, where 
$r_h = M/v_0^2$, $M=4 \times 10^6~{\rm M}_{\odot}$~\cite{GheSWLD08,GenEG10}  
and $v_0 = 105~{\rm km s^{-1}}$~\cite{GulRGLT09}. The 
inner boundary is at $r_{\rm bh} = 6 \times 10^6~{\rm km}$. 
From the DM density in the solar neighborhood, 
$\rho_D = 0.008 ~{\rm M}_{\odot} {\rm pc^{-3}}$~\cite{BovT12} 
at a distance $D=8.5~{\rm kpc}$ from the Galactic center~\cite{DayFHLPRS16}, 
we infer the DM 
density at $r_b$ to be $\rho_b = \rho_D (D/r_b)^{\gamma_c} = 
2.8 \times 10^3~{\rm M}_{\odot} {\rm pc}^{-3}$.

The DM annihilation cross sections are given by
\begin{equation}
\label{cross}
\langle \sigma v \rangle = {\langle \sigma v \rangle}_{\rm can} 
\left( \frac{v^2}{v_{\rm fo}^2} \right)^s
\end{equation}
where $s=0$ for $s$-wave annihilation and $s=1$ for $p$-wave annihilation.
Here we follow~\cite{DayFHLPRS16,FieSS14} and take ${\langle \sigma v \rangle}_{\rm can} = 
1.7 \times 10^{-26}~{\rm cm}^3 {\rm s}^{-1}$,
close to the value expected for a thermal relic origin of DM, with 
the freeze-out parameter $v_{\rm fo}=c/4$ for $s=1$. For the DM mass we
choose $m_{\chi}=35$~GeV.

Given the above particle models we calculate 
that at $t = T=10^{10}~{\rm yr}$ the annihilation plateau densities defined by 
Eqn.~(\ref{rhoann})
in the DM spike are  
$\rho_{\rm ann}(s{\rm-wave}) =1.7 \times 10^8 ~{\rm M}_{\odot} {\rm pc}^{-3}$ 
and $\rho_{\rm ann}(p{\rm-wave})=
6.6 \times 10^{10} ~{\rm M}_{\odot} {\rm pc}^{-3}$. 
These densities are reached at radii $r_{\rm ann}(s{\rm-wave}) = 
3.1 \times 10^{-3}~{\rm pc}$ and $r_{\rm ann}(p{\rm-wave}) = 2.5 \times 10^{-4}~{\rm pc}$
in the spike, within which we expect the density spike to 
transition to a weak cusp.
The cusp is smaller for $p$-wave than for $s$-wave annihilation since the 
annihilation cross section is reduced by $\sim v^2/c^2$, so the time scale for 
$p$-wave annihilation to destroy matter in the innermost spike 
is correspondingly longer.
 
The Boltzmann equation may be written as 
\begin{equation}
\frac{\partial f({\bf r},{\bf v};t)}{\partial t} = 
-\frac{\rho({\bf r})}{m_{\chi}} 
\langle \sigma v \rangle  f({\bf r},{\bf v};t),
\end{equation}
which can be transformed to yield
\begin{equation}
\frac{\partial f(E,R;t)}{\partial t} = 
-\frac{\overline {\rho \langle \sigma v \rangle}}{m_{\chi}} f(E,R;t),
\end{equation}
or
\begin{equation}
\label{boltz}
\frac{\partial f(E,R;\tau)}{\partial \tau} =
- \frac{\overline{\rho v^{2s}}}{\rho_{\rm a} v_{\rm fo}^{2s}} f(E,R;\tau).
\end{equation}
Here  $\tau = t/T$,  $\rho_a$ is given by Eqn.~(\ref{rhoann}) 
for $s=0$, and is thus a constant, 
and the overbar denotes a radial average over orbital period $P(E)$,
\begin{eqnarray}
\label{avg}
\overline{\rho v^{2s}}&=&\frac{1}{P(E)}
\oint {\rho(r) v^{2s}(r)}\frac{dr}{v_{\hat{r}}} \\
&=& \int_{1-\sqrt{1-R}}^{1+\sqrt{1-R}} \rho(xr_c) v^{2s}(xr_c) 
\frac{dx}{\pi \sqrt{2/x - 1 - R/x^2}}. \nonumber
\end{eqnarray}

In writing Eqn.~(\ref{avg}) we set $r = xr_c$, where $r_c = M/(-2E)$ is the radius
of a circular orbit with energy $E$. The profiles for $\rho$ and $v$ appearing in
the integrands in  Eqn.~(\ref{avg}) are obtained at time $\tau$ from  
Eqn~(\ref{moments}), with $f_{\rm loss}(E,R)$ replaced by the current value
of $f(E,R;\tau)$. Loss-cone boundary conditions are imposed
throughout the evolution.  
We take as initial data an adiabatic distribution function specified
by Eqn.~(\ref{fER}), with $f(E) = K|E|^p, p=0.86$ and $r_{\rm ann}=0$
(i.e., no annihilation imprint at $\tau$ =0).

\begin{figure}
\includegraphics[width=8cm]{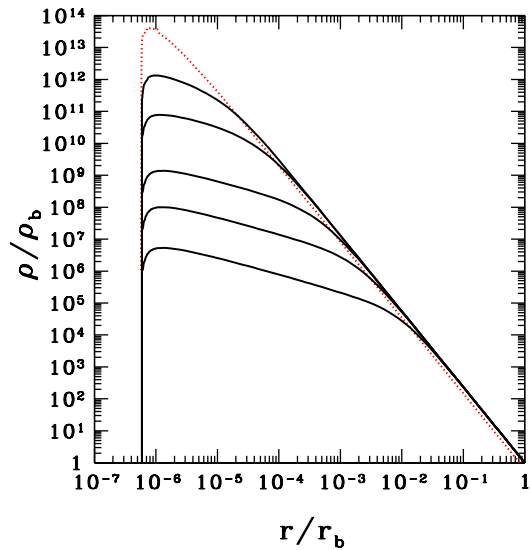}
\includegraphics[width=8cm]{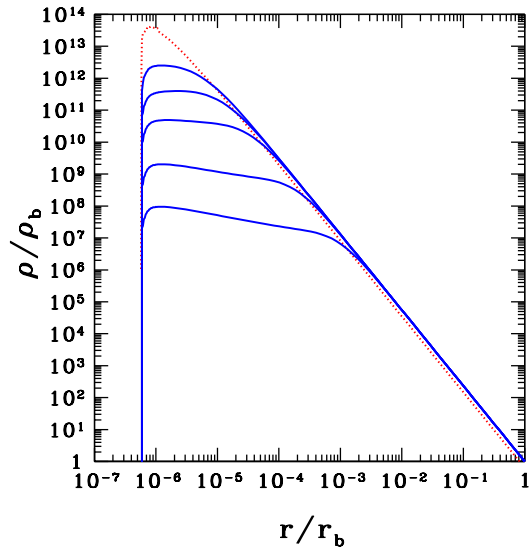}
\caption{Evolution of the density profile in a 
DM spike around Sgr A*, allowing for $s$-wave ({\it top}) and 
$p$-wave ({\it bottom}) annihilation and black hole capture.
The {\it dotted} curve shows the initial adiabatic profile at $t=0$.
Moving downward, successive {\it solid} curves show the profiles at 
$t/T = 1.6 \times 10^{-7}, ~4.8 \times 10^{-6}, ~8.2 \times 10^{-4}, 
~2.4 \times 10^{-2}$ and $1.0$ ({\it top}) and at $t/T = 4.9 \times 10^{-8}, 
~7.6 \times 10^{-7}, ~2.0 \times 10^{-5}, ~5.1 \times 10^{-3}$ and 
$1.0$ ({\it bottom}), where $T = 10^{10}~$yr.
The densities and radii are normalized to their values near the 
spike outer boundary at $r_b = 0.34~$pc, 
where $\rho_b=2.8 \times 10^3~{\rm M}_{\odot} {\rm pc}^{-3}$. 
}
\label{fig:dens_MW}
\end{figure}

\begin{figure}
\includegraphics[width=8cm]{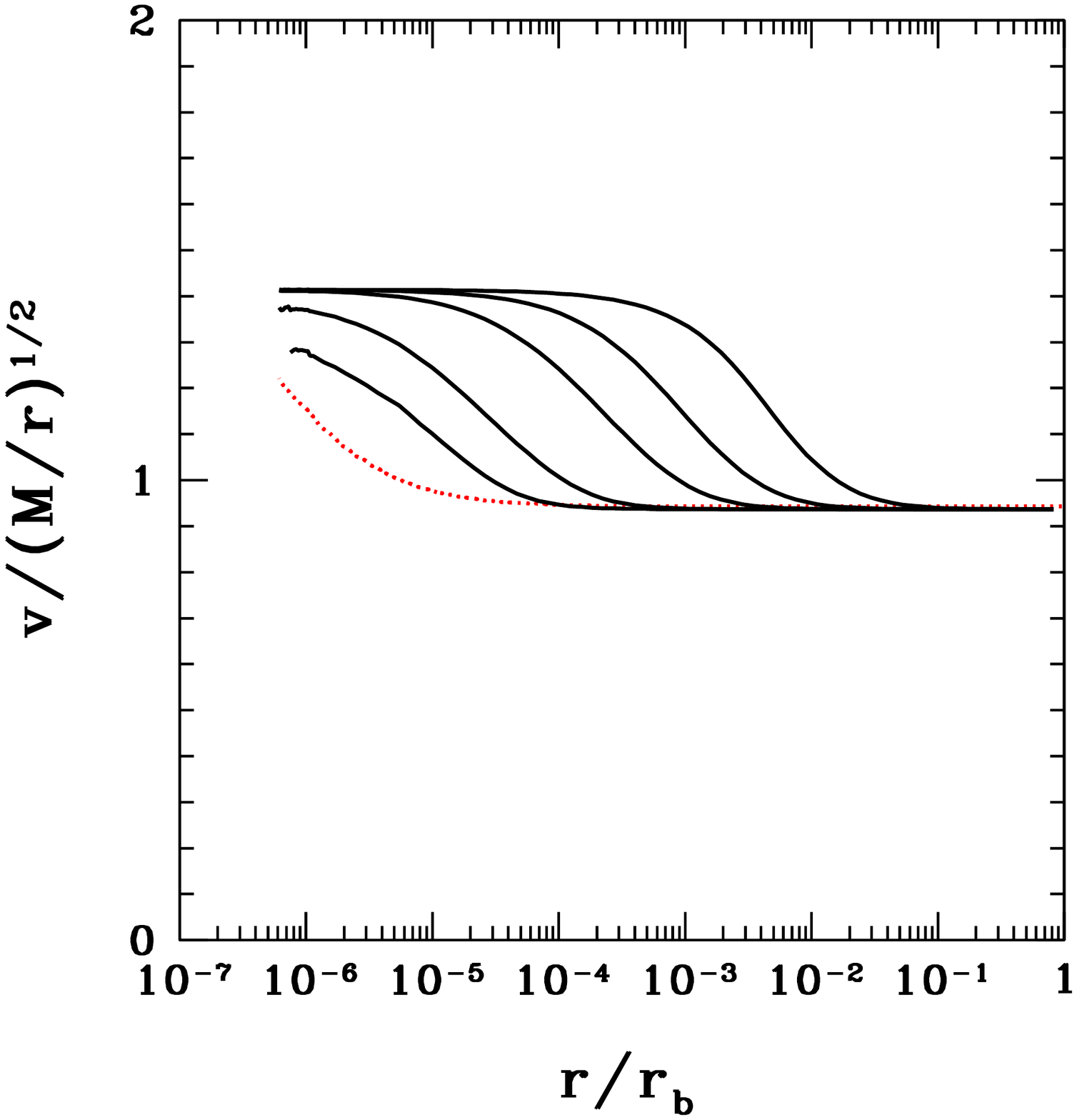}
\includegraphics[width=8cm]{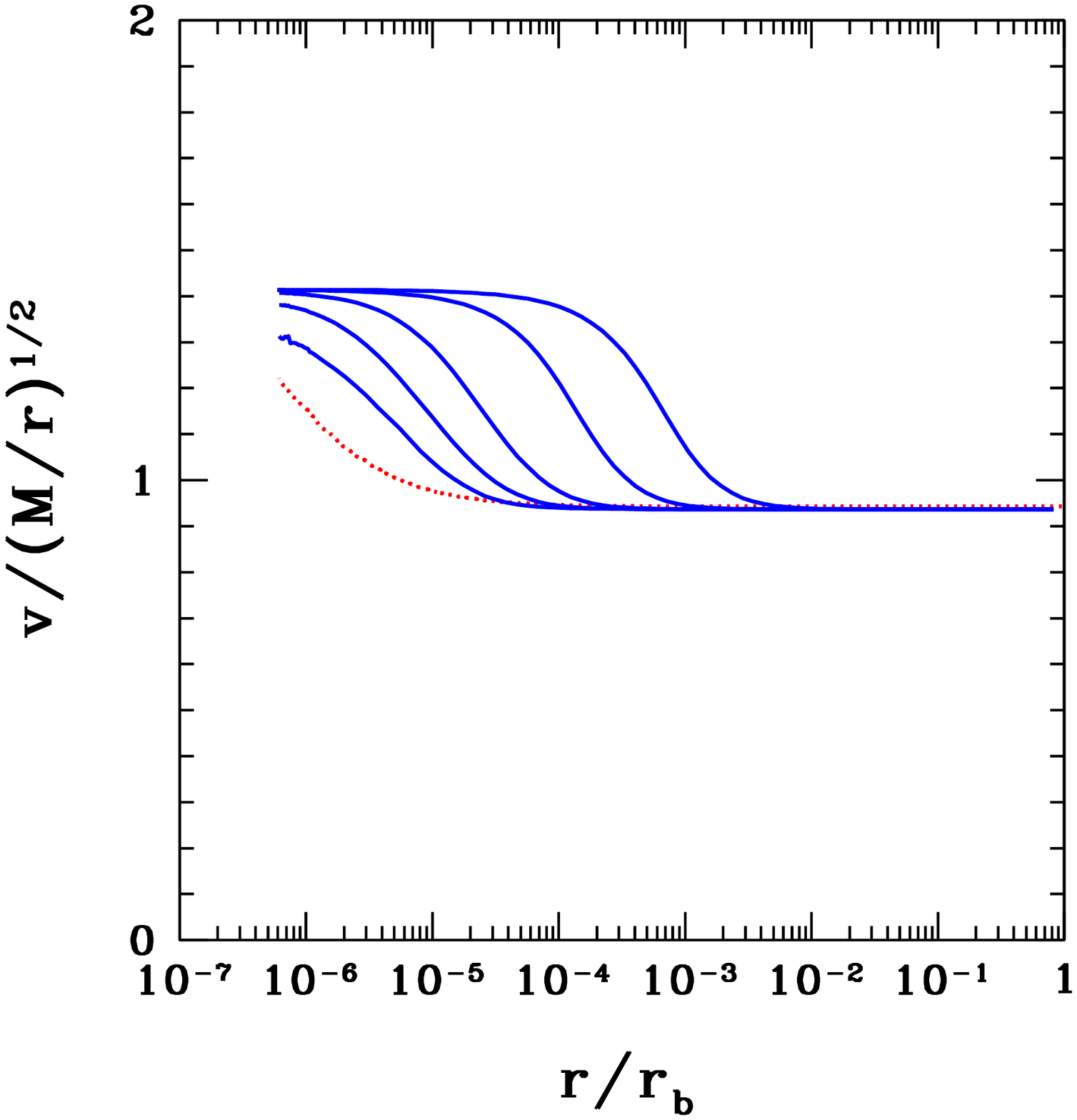}
\caption{Evolution of the velocity dispersion (3D) in a
DM spike around Sgr A*, allowing for $s$-wave ({\it top}) and 
$p$-wave ({\it bottom}) annihilation and black hole capture.
The {\it dotted} curve shows the initial adiabatic profile at $t=0$. 
Moving upward, successive {\it solid} curves show the profiles at
$t/T = 1.6 \times 10^{-7}, ~4.8 \times 10^{-6}, ~8.2 \times 10^{-4}, 
~2.4 \times 10^{-2}$ 
and $1.0$ ({\it top}), and at $t/T = 4.9 \times 10^{-8}, ~7.6 \times 10^{-7}, 
~2.0 \times 10^{-5}, ~5.1 \times 10^{-3}$ and $1.0$  ({\it bottom}),
where $T = 10^{10}~$yr. 
Radii are normalized to the value near the spike outer 
boundary at $r_b = 0.34~$pc, 
and velocities are normalized to $(M/r)^{1/2}$.
}
\label{fig:vel_MW}
\end{figure}

We integrate the evolution Eqn.~(\ref{boltz}) by finite differencing in $E$ and $R$ and
evolving in time $\tau$ by a first-order semi-implicit method.
All time integrations and 
phase-space quadratures are repeated with finer resolution to check reliability.
Results for the density and velocity profiles are summarized in 
Figs.~(\ref{fig:dens_MW}) and (\ref{fig:vel_MW}), respectively. 
 
The $s$-wave profile in Fig.~(\ref{fig:dens_MW}) exhibits a weak cusp
inside the annihilation region at each time, within which the density varies 
as $r^{-1/2}$. This result is in accord with our simplified models 
constructed in Sections ~\ref{iso} and \ref{lossc}. 
As $\rho_{\rm ann}$ decreases with time, the weak cusp grows, 
eating its way outward into the steeper spike.
The $p$-wave profile behaves qualitatively similarly, with two
notable differences. The first is that for the same evolution time
the $p$-wave cusp is smaller, as described above. The second is
that the $p$-wave cusp is somewhat shallower, varying as $r^{-0.34}$
rather than $r^{-1/2}$. This may be understood by noting that
the annihilation plateau density $\rho_{\rm ann}$ given by 
Eqn.~(\ref{rhoann}) decreases with decreasing distance from the black
hole, since the velocity dispersion and annihilation cross section
increase. Hence while the cusp is still filled with high eccentricity
particles from outside the cusp that plunge inside at pericenter, the
lower eccentricity particles in the cusp are driven to lower 
(``plateau'') densities the closer they are to the black hole.
This effect causes the overall slope of the density profile in the cusp to
fall slightly below $1/2$ to $\sim 0.34$ by $t=10^{10}$~yr.

The velocity profiles plotted in Fig.~\ref{fig:vel_MW} also show
that the cusps grow in size with time and at any one time 
are larger for $s$-wave annihilation than for
$p$-wave annihilation.  Otherwise the profiles in 
are identical in the unperturbed spike regions and very close in the
cusp regions, conforming to those found for the simplified models in 
Section ~\ref{iso} and \ref{lossc}.

\begin{figure}
\includegraphics[width=8cm]{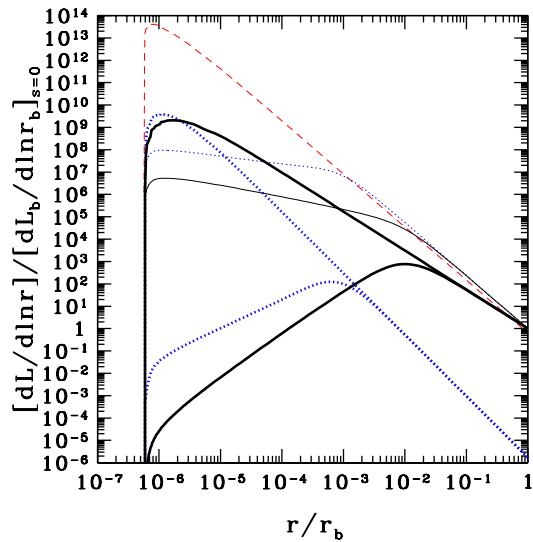}
\caption{The luminosity profile from annihilation in a DM
spike around Sgr A*. The {\it heavy solid} ({\it black}) 
curves show the luminosity
for $s$-wave annihilation at $t=0$ ({\it upper}) and 
at $t = 10^{10}~{\rm yr}$ ({\it lower}). 
The {\it heavy dotted} ({\it blue}) curves show
the luminosity for $p$-wave annihilation at $t=0$ ({\it upper}) and 
at $t = 10^{10}~{\rm yr}$ ({\it lower}). 
For comparison, the {\it dashed} ({\it red}) curve 
shows the DM adiabatic density profile at $t=0$, while the 
density profile at $t = 10^{10}~{\rm yr}$ is 
shown for $s$-wave annihilation by the {\it thin solid} ({\it black}) curve
and for $p$-wave annihilations by the {\it thin dotted} ({\it blue}) curve. 
All luminosities are normalized by the initial 
$s$-wave luminosity at the spike outer boundary at $r_b = 0.34~$pc. 
All radii are normalized by $r_b$.
}
\label{fig:lum_MW}
\end{figure}

Next we consider the luminosity profiles arising from DM annihilation within
the spike. The photon luminosity emerging from radius $r$ is given by
\begin{equation}
\label{lum}
L(r) = \int_{r_{\rm bh}}^r \frac{1}{2} \frac{\rho(r)^2}{m_{\chi}^2} 
(2\epsilon_{\gamma} m_{\chi}) \langle \sigma v \rangle 4 \pi r^2 dr,
\end{equation} 
where $\epsilon_{\gamma}$ is the fraction of the annihilation energy
that goes into photons.
The region between $r$ and $2r$ that contributes most of the luminosity is 
centered near the peak of the function $dL(r)/d~ln(r)$, where
according to Eqns.~(\ref{cross}) and (\ref{lum}),
\begin{equation}
\frac{dL(r)}{d~ln(r)} \propto r^3 \rho^2 
\left( \frac{v^2(r)}{v_{\rm fo}^2} \right)^s.  
\end{equation}
This function is plotted in Fig.~\ref{fig:lum_MW} for the two cases,
along with the corresponding density profiles. Results are shown for
both the initial spike and the spike at $t=T=10^{10}$~yr.  Several
features are evident from the plot. The first is that for both
$s$-wave and $p$-wave annihilation the dominant emission originates
from the innermost region of the spike near $r \gtrsim r_{\rm bh}$
initially, but moves out to the outer edge of the weak cusp $r \sim
r_{\rm ann}$ at later times. As annihilations eat their way further
into the spike and $r_{\rm ann}$ moves outward with time, the
magnitude of the luminosity falls.  Apart from the initial time, when
the luminosities are comparable, the luminosity is greater for
$s$-wave annihilation than for $p$-wave annihilation. This difference
results from the fact that the main radiating region around 
$r_{\rm ann}$ has a much smaller volume and the cross section has an
additional factor of $v^2/c^2$ for $p$-wave versus $s$-wave
annihilation. 

We note that for a flat plateau instead
of a weak cusp the luminosity profile plotted in Fig.~\ref{fig:lum_MW} 
would plummet faster for all 
$r< r_{\rm ann}$ and thereby reduce the overall annihilation flux.
For the Galactic parameters adopted here it is a $\sim 10\%$ reduction
for $s$-wave annihilation and less for $p$-wave annihilation, but
can be larger for different parameters or DM halos. 

Fig.~\ref{fig:lum_MW} shows that most of the luminosity
from the spike originates from the region around $r_{\rm ann}$ and
that $r_{\rm ann} \gg M$ at $t=10^{10}$~yr. As a result, our Newtonian
analysis of the bulk profiles in this region and, hence, the annihilation
luminosity, are little modified by relativistic corrections. However, it
has been suggested that a high-energy tail in the (gamma-ray) spectrum
might arise from the Penrose process in the vicinity of a rapidly
spinning Kerr black hole~\cite{Sch15}. Here a fully relativistic
treatment is necessary, but the ambient spike and weak cusp should be
close to the profiles obtained here for all $r \gg M$. 

\section{Summary}

We have reinvestigated the effect
of DM self-annihilations on the distribution of collisionless DM in a
spherical density spike around a BH.  These spikes can reach the
so-called ``annihilation plateau'' density $\rho_{\rm ann} =
m_\chi/(\langle\sigma v \rangle T)$ at a radius $r=r_{\rm ann}$, where
the time scale for DM annihilation becomes equal to the age of the
Galaxy.  Interior to this radius, DM annihilations are important for
determining the radial density and velocity dispersion profiles of DM,
with potentially observable consequences for indirect detection.  We
revisit and extend the results of \cite{Vas7} for $s$-wave
annihilation cross sections, and provide the first results for
nonconstant annihilation cross sections, with the very well-motivated
case of $p$-wave annihilations.

We first give a simple physical argument for the case of an isotropic
phase space distribution function that yields analytic expressions
for the DM density and velocity dispersion profiles within a DM spike
with a weak cusp.  This argument reproduces the result of \cite{Vas7}
for the DM density profile in the case of a velocity-independent
$s$-wave annihilation cross section, where the density follows a power
law $\rho(r)\propto r^{-1/2}$ for radii below $r_{\rm ann}$.  We then
extend this analytic model to incorporate the direct capture of DM
particles by the BH via a loss-cone boundary condition, making the
resulting distribution anisotropic.  Finally, to provide a full
description of the (spherically symmetric) system, we integrate the
collisionless Boltzmann equation numerically and study the formation
of the weak cusp and its subsequent evolution with time.  We find that
the increasing annihilation cross section at decreasing radii in the case
of $p$-wave annihilations flattens the annihilation cusp relative to
that obtained with $s$-wave
annihilations, yielding $\rho(r)\propto r^{-0.34}$ for
the Galactic parameters adopted here, but still yields a cusp.

\medskip 

{\it Acknowledgments}: It is a pleasure to thank B. Fields for helpful
discussions. This paper was supported in part by NSF Grant No. PHY-1300903
and NASA Grant No. NN13AH44G at the University of Illinois at
Urbana-Champaign.

\bibliography{paper}
\end{document}